\documentclass[preprint,fleqn,showpacs,showkeys, superscriptaddress]{revtex4}
\usepackage{graphicx}
\usepackage{amssymb}
\usepackage{amsmath}
\usepackage{bm}

\begin{document}

\title{Statistical Investigation of Connected Structures of Stock Networks in Financial Time Series}

\author{Cheoljun Eom}
\email{shunter@pusan.ac.kr} \affiliation{Division of Business
Administration, Pusan National University, Busan 609-735, Korea}

\author{Gabjin Oh}
\affiliation{NCSL, Department of Physics, Pohang University of
Science and Technology, Pohang, Gyeongbuk, 790-784, Korea}

\author{Seunghwan Kim}
\affiliation{NCSL, Department of Physics, Pohang University of
Science and Technology, Pohang, Gyeongbuk, 790-784, Korea}

\begin{abstract}
In this study, we have investigated factors of determination which
can affect the connected structure of a stock network. The
representative index for topological properties of a stock network
is the number of links with other stocks. We used the multi-factor
model, extensively acknowledged in financial literature. In the
multi-factor model, common factors  act as independent variables
while returns of individual stocks act as dependent variables. We
calculated the coefficient of determination, which represents the
measurement value of the degree in which dependent variables are
explained by independent variables. Therefore, we investigated the
relationship between the number of links in the stock network and
the coefficient of determination in the multi-factor model. We
used individual stocks traded on the market indices of Korea,
Japan, Canada, Italy and the UK. The results are as follows. We
found that the mean coefficient of determination of stocks with a
large number of links have higher values than those with a small
number of links with other stocks. These results suggest that
common factors are significantly deterministic factors to be taken
into account when making a stock network. Furthermore, stocks with
a large number of links to other stocks can be more affected by
common factors.
\end{abstract}

 \maketitle

\section{Introduction}

Stock markets are known to form a pricing mechanism by
interactions of stocks having a variety of characters. The
academic world recognizes that one of its major research tasks is
to observe the dynamics of such interactions between stocks.
Recently, stock networks have been throughly analyzed in the field
of econophysics to understand the interactions between stocks.
Mantegna visually presented the complicated relationships between
stocks in a topological space using the minimal spanning tree
(MST) method. He achieved this by taking advantage of the degree
of correlations between stocks, which is called a stock network
[1]. Thereafter, many studies have verified diverse properties of
stock networks derived from the MST method by using time series
data of financial markets of various countries [2-7]. In the
meantime, there have also been studies which have observed stock
networks derived from the same test process using returns
generated directly from theoretical pricing models [8-10]. These
studies were under taken in order to observe the formation
principles and factors of determination for the connected
structures of stock networks.

Based on statistics, we examined factors of determination which
can affect the link structures of stock networks. A major index
which represents a link structure of a stock network is the number
of links which a specific stock has with other stocks. That is,
most stocks have a small number of links with other stocks but a
few stocks have a very large number of links with other stocks.
Exploring the factors of determination that can influence the
number of links of these stocks will help us to understand the
formation principles and factors of determination for the
connected structures of stock networks. Of the factors of
determination available, we use common factors which are widely
recognized in the field of finance. Common factors refer to
factors that can commonly influence stocks and major common
factors mentioned in the finance sector. These include market
factors, industrial factors, macroeconomic factors and corporate
factors [11-13]. Examples of major pricing models reflecting such
common factors are the one-factor model [14], three-factor model
[15] and multi-factor model [16]. We use the multi-factor model,
which is generated by a statistical approach, to verify the
connected structures of stock networks. According to the results
of our observation, stocks which have a large number of links to
other stocks in stock networks are more likely to be explained by
common factors than those which have a small number of links. This
indicates the fact that common factors are the factors of
determination that can have a significant influence on the link
structures of stock networks.

In the next section, we describe the data and methods of the test
procedures used in this paper. In section III, we present the
results obtained according to our established research aims.
Finally, we summarize the findings and conclusions of the study.

\section{DATA and METHODS}

\subsection{Data}

We used the individual stocks traded on the stock market indices
of Korea, Japan, Canada, Italy, and the UK. That is, we used the
daily prices of 127 stocks in the KOSPI 200 market index of the
Korean stock market, 202 stocks in the Nikkei 225 of Japan, 118
stocks in the TSX of Canada, 111 stocks in the Milan Comit General
of Italy, and 69 stocks in the FTSE 100 of the UK. The individual
stocks that had daily prices for the last 15 years, from January
1992 to December 2006, were selected from each country. The
returns, $R(t)$, are calculated by the logarithmic change of the
price, ${R(t)} = \ln{P(t)}- \ln{P(t-1)}$, where $P(t)$ is the
stock price at $t$ day.

\subsection{Minimal Spanning Tree Method}

The stock network visually displays the significant $N-1$ links
among all possible links, $N(N-1)/2$. This is based on the
correlation matrix between stocks, using the MST method. The MST,
a theoretical concept in graph theory [17], is also known as the
single linkage method of cluster analysis in multivariate
statistics [18-19]. We created the stock networks using the
correlation matrix calculated from the return, $R_j$. The
correlation matrix, $\rho_{i,j}$, is defined by

\begin{gather}
\rho_{i,j} \equiv \frac{\langle R_i R_j \rangle - \langle R_i
\rangle \langle R_j \rangle} {\sqrt{(\langle R_{i}^{2}\rangle -
\langle R_{i}\rangle ^2) (\langle R_{j}^{2}\rangle - \langle
R_{j}\rangle ^2) }}, \tag{1a}
\end{gather}

where the notation $\langle \cdots \rangle$ means an average over
time. In order to create the stock network, the metric distance,
$d_{i,j}$, relates the distance between two stocks to their
correlation coefficient [20], and is defined as

\begin{gather}
d_{i,j} = \sqrt{2(1-\rho_{i,j})}, \tag{1b}
\end{gather}

The MST is the spanning tree of the shortest length using the
Kruskal or Prim algorithm [21-22]. In our study, we used the
Kruskal algorithm. The Kruskal algorithm is a graph without a
cycle connecting all nodes with links. The correlation coefficient
can vary between $-1 \leq \rho_{i,j} \leq +1$ while the distance
can vary between $0 \leq d_{i,j} \leq 2$. Here, small values of
the distance imply strong correlations between stocks.

In this paper, using the  individual stocks traded on the stock
market index of each country, we created a stock network by
utilizing the MST method. The largest number of links with other
stocks in the stock networks of each country is 14 for Korea, 15
for Japan, 12 for Canada, 16 for Italy and 10 for the UK. In
addition, we needed to assure standardization by the maximum and
minimum links because the degree distribution of stock networks
for each country is different. According to the results, we found
that the degree distribution of stock networks follows a power-law
with the exponent $\overline{\gamma} \approx -2.36$ . In other
words, most stocks have a small number of links to other stocks,
while a few stocks have a large number of links to other stocks.

\subsection{Multi-factor Model}

In this paragraph, we look into the test process of the
multi-factor model, a theoretical concept of the arbitrage pricing
model [23]. The independent variables used in the multi-factor
model are common factors that are estimated through factor
analysis in multivariate statistics [24]. Factor analysis widely
used in the field of social science can reduce the many variables
in the given data set to a few factors. Using factor analysis, we
chose significant factors which are regarded as having economic
significance. Accordingly, we regard significant factors as common
factors. We also created the new time series having the attributes
of significant factors, called factor scores in statistics. That
is, the factor scores are time series data of common factors.
Furthermore, the factor scores are used as independent variables
in the multi-factor model. In the multi-factor model, the stock
returns, $R_{j}(t), ~ j=1,2,...,N$, can be explained by the common
factors, $F_{k}(t), ~ k=1,2,...,K$. They are defined by

\begin{gather}
R_{j}(t) = \alpha_{j} + \beta_{j,1}F_{1}(t) + ...... +
\beta_{j,k}F_{k}(t) + \epsilon_{j}(t),\tag{2}
\end{gather}

where $\alpha_{j}$ is an expected return on the stock,
$\beta_{j,k}$ are a stock's sensitivity to changes in common
factors, and $\epsilon_{j}(t)$ is the residuals [$E(\epsilon_{j})
\approx 0$, $E(\epsilon_{j}, \epsilon_{m}) \approx 0$, and
$E(\epsilon_{j}, F_{k}) \approx 0$]. To establish the multi-factor
model as in Eq. 2, we have to determine the number of common
factors, $K$, and control the problem of multicolinearity between
common factors [25]. First, in order to determine the number of
significant factors, we used the Kaiser rule [26]. That is,
measured eigenvalues determine the factors corresponding to the
number having a value of "1" or higher as the number of
significant factors. Second, results exhibiting the problem of
multicolinearity might be distorted according to the higher
correlation between the independent variables. Therefore, in order
to minimize the correlation between common factors, we created a
new time series according to factor analysis controlled by the
rotated varimax method.

In this paper, the number of common factors chosen by the Kaiser
rule is 10 for Korea, 20 for Japan, 7 for Canada, 7 for Italy, and
8 for the UK. The correlation between the common factors have much
lower values and the mean value of the correlations is $2.83\%$
for Korea, $1.11\%$ for Japan, $4.79\%$ for Canada, $3.19\%$ for
Italy, and $3.56\%$ for the UK. That is, there are no problems
concerning multicolinearity that may occur in multi-factor models.

\section{Results}

In this section, we present the observed results. The
representative index for topological properties of a stock network
is the number of links with other stocks. In the multi-factor
model, we can evaluate whether common factors such as independent
variables are the important deterministic factors according to
whether the independent variables can explain a dependent
variable. That is, the major index is the coefficient of
determination, $ 0\% \leq R^{2} \leq 100\%$. The coefficient of
determination has a higher value if the independent variables
sufficiently explain a dependent variable. Therefore, we
investigated the relationship between the number of links in the
stock network and the coefficient of determination in the
multi-factor model.

\begin{figure}[tb]

\includegraphics[height=7cm, width=15cm]{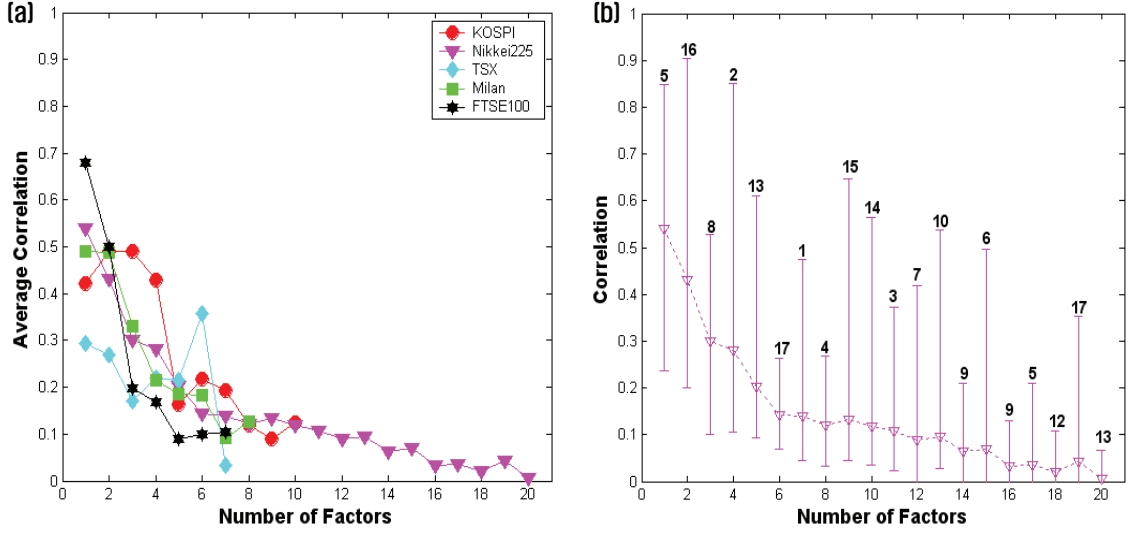}

\caption[0]{The figure indicates the correlations between the
common factors and the industrial average indexes. In Fig. 1(a),
axis x represents the numbers of common factors, and axis y
indicates correlations between the common factors and the
industrial average indexes. In the figure, Korea is indicated as
red circles, Japan as magenta triangles, Canada as cyan diamonds,
Italy as green boxes, and the UK as black pentagrams. In fig.
1(b), the maximum, average and minimum value of the correlation
matrix calculated for Japan are indicated as an error-bar graph.
The number of the industrial average indexes used is 17 ( 1:
Automobiles and Parts, 2: Banks, 3: Chemicals, 4: Construct. and
Material, 5: Food Producers, 6: General Financial, 7: General
Retailers, 8: Industrial Engineering, 9: Industrial Metals, 10:
Industrial Transport, 11: Leisure Goods, 12: Personal Goods, 13:
Pharm. and Biotech, 14: Software and Comp. Svs, 15: Support
Services, 16: Tech Hware and Equipment, and 17: Travel and Leisure
). The numbers indicated on the top of the graph in the figure
represent the numbers of the industrial average indexes which have
the maximum value of correlations for each common factor.}

\end{figure}

First of all, we examined the economic significance of common
factors used as independent variables in the multi-factor model.
Common factors mentioned in the field of finance include market
effects, industrial effects and macroeconomic effects. We
calculated measurement values that can show the properties of
industrial effects, $\overline{R_t^{i}} = \frac{1}{N_i}
\sum_{j=1}^{N_{i}} R_{j,t}^i$. This represents the average value
of $R_{j,t}^i$ which is the returns of $N_i$ individual stocks
belonging to a specific industry, $i$, that was defined as an
industrial average index having an industrial property. That is,
this corresponds to methods that calculate the equal-weighted
indices which are widely used in the field of finance. To
calculate a measurement value having industrial effects
properties, however, we excluded the industries that have four or
less individual stocks. The number of industrial average indexes
obtained through this calculation process was 10 for Korea, 17 for
Japan, 7 for Canada, 8 for Italy, and 2 for the UK. The results
are shown in Fig. 1.

Figure 1 indicates the correlations between common factors and
industrial average indexes. In Fig. 1(a), axis x represents the
number of common factors. The highest number of common factors is
20 for Japan; therefore, axis x had a maximum value of 20.
Correlations between the common factors and the industrial average
indexes are calculated and the average values of those
correlations are indicated on axis y. For Japan, for example, the
number of common factors is 20 and the number of industrial
average indexes is 17. The average correlations between the first
common factor and the 17 industrial average indexes are indicated.
Similarly, the average correlation coefficients from the second
common factor to the last common factor are repeatedly indicated.
In Figure 1, Korea is indicated by red circles, Japan by magenta
triangles, Canada by cyan diamonds, Italy by green boxes, and the
UK by black pentagrams. For Fig. 1(a), the maximum, average and
minimum value of the correlation matrix calculated for Japan are
indicated by an error-bar graph in Fig. 1(b). The number of
industrial average indexes used is 17. The numbers indicated on
the top of the graph in the figure represent the numbers of the
industrial average indexes which have the maximum value of
correlations for each common factor.

According to the results, the correlations between common factors
and industrial average indexes are high on average within the
range of common factors 1-3, regardless of the data used. However,
the degree of correlations rapidly declines thereafter in Fig.
1(a). In addition, each common factor has a high correlation with
a specific industrial average index in Fig. 1(b). That is, most of
the 20 common factors selected from the Japanese stock market have
the highest correlation with different industrial average indexes
among the 17 industrial average indexes. These results indicate
that, although time series data derived from statistics were used
as independent variables of the multi-factor model, these data
have the same economic significance as the common factors
mentioned in the field of finance.

\begin{figure}[tb]

\includegraphics[height=7cm, width=15cm]{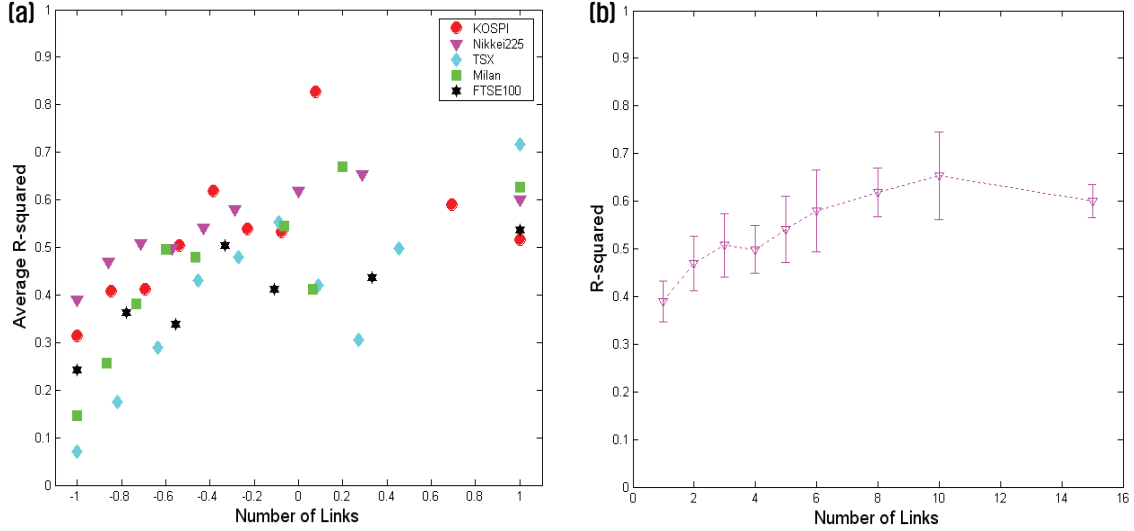}

\caption[0]{The figure indicates the relationship between the
number of links in a stock network and the average coefficient of
determination in the multi-factor model. In Fig. 2(a), axis x
indicates the number of links. The axis x denotes the normalized
connecting number from -1 to +1 using the minimum and maximum
connecting number of each country. Axis y represents the average
coefficient of determination of individual stocks belonging to
each number of links. In the figure, Korea is indicated as red
circles, Japan as magenta triangles, Canada as cyan diamonds,
Italy as green boxes and the UK as black pentagrams. Fig. 2(b)
indicates an error-bar graph of the maximum, average and minimum
value of the coefficients of determination for each number of
links for Japan. The maximum number of links in a stock network is
15 for Japan.}

\end{figure}

Next, we investigated the relationship between common factors and
 link structures of stock networks. In the multi-factor model,
common factors are independent variables while returns of
individual stocks are dependent variables. We calculated the
coefficient of determination, $R^2$, which represents the
measurement value of the degree in which dependent variables are
explained by independent variables. Then, we investigated the
number of links, $L(k)$, $k=1,2,...,Max$, that each stock has with
other stocks in a stock network by utilizing the MST method. Then,
we calculated an average coefficient of determination of
individual stocks, $\overline{R_{L(k)}^{2}}$, which are included
in each number of links. The results are shown in Fig. 2.

Figure 2 indicates the relationship between the number of links in
a stock network and the average coefficient of determination in
the multi-factor model. In Fig. 2(a), x-axis indicates the number
of links. The largest number of links to other stocks in the stock
networks of each country is 14 for Korea, 15 for Japan, 12 for
Canada, 16 for Italy and 10 for the UK. In addition, we needed to
assure standardization
$[L^{*}=(L_{i}-L_{Min})/(L_{Max}-L_{Min})-1]$ by the maximum
$(L_{Max})$ and minimum $(L_{Min})$ links, because the degree
distribution of stock networks for each country is different. The
normalized value $L^{*}$ lies within $-1 \leq L^{*} \leq +1$. Axis
y represents the average coefficient of determination of
individual stocks belonging to each number of links. In Fig. 2,
Korea is indicated by red circles, Japan by magenta triangles,
Canada by cyan diamonds, Italy by green boxes and the UK by black
pentagrams. Meanwhile, regarding the results of Fig. 2(a), Fig.
2(b) indicates an error-bar graph of the maximum, average and
minimum value of the coefficients of determination for each number
of links for Japan. The maximum number of links in a stock network
is 15 for Japan.

The results indicate that stocks having more links to other stocks
in a stock network have higher average coefficients of
determination. That is, when the common factors were used as
independent variables of the multi-factor model, stocks with more
links to other stocks show higher coefficients of determination
that measure the degree of explanation of changes in returns of
individual stocks. Compared with stocks with less links to other
stocks in a stock network, stocks with more links are more likely
to be explained by common factors. The results suggest that common
factors are the factors of determination that can have a
significant influence on the formation process of connected
structures of stock networks.

\section{Conclusions}

We investigated factors of determination that can affect link
structures of stock networks in viewpoint of statistics. The
number of links which a specific stock has with other stocks was
used as a major index that represents the connected structure of a
stock network. Of possible factors of determination, common
factors widely recognized in the field of finance were used. Thus,
exploring the factors of determination that can influence the
number of links of a stock is expected to be of great help in
understanding formation principles and factors of determination
for link structures of stock networks. For a pricing model
reflecting common factors, we verified link structures of stock
networks by using a multi-factor model generated by a statistical
approach.

Our study produced the following conclusions. First, we confirmed
the economic significance of common factors which were used as
independent variables in the multi-factor model. We found that
common factors extracted by a statistical base have industrial
effects properties mentioned in the field of finance. That is,
common factors have an economic significance. Second, we verified
the relationship between common factors and the number of links
which represents the link structure properties of a stock network.
We discovered that stocks having more links to other stocks in a
stock network are more likely to be explained by common factors
than those with fewer links. Through these results, we have
demonstrated that the common factors mentioned in the field of
finance are factors of determination that can have a significant
influence on the formation process of link structures of stock
networks.


\begin{thebibliography}{00}

\bibitem{1}
R.N.Mantegna, Eur. Phys. J. B 11 (1999) 193.

\bibitem{2}
G.Bonanno, N.Vandewalle, and R.N.Mantegna, Phys. Rev. E 62(6)
(2000) 7615.

\bibitem{3}
G.Bonanno, F.Lillo, and R.N.Mantegna, Quant. Finance 1 (2001) 96.

\bibitem{4}
G.Bonanno, G.Caldarelli, F.Lillo, S.Micciche, N.Vandewalle, and
R.N.Mantegna, Eur. Phys. J. B 38 (2004) 363.

\bibitem{5}
R.Coelho, S.Hutzler, P.Repetowicz, and P.Richmond, Physica A,
373(1) (2007) 615.

\bibitem{6}
J.P.Onnela, A.Chakraborti, K.Kaski, J.Kertesz, and A.Kanto, Phys.
Rev. E 68(5) (2003) 056110.

\bibitem{7}
J.P.Onnela, A.Chakraborti, K.Kaski, and J.Kertesz, Physica A 324
(2003) 247.

\bibitem{8}
G.Bonanno, G.Caldarelli, F.Lillo and R.N.Mantegna, Phys. Rev. E 68
(2003) 046130.

\bibitem{9}
Cheoljun Eom, Gabjin Oh, and Seunghwan Kim, Physica A 383 (2007)
139.

\bibitem{10}
Cheoljun Eom, Gabjin Oh, and Seunghwan Kim, Preprint available at
arxiv.org, physics/0612068, 2006.

\bibitem{11}
B.F.King, J. Business 39(1) (1996) 139.

\bibitem{12}
J.L.Farrell, Jr., J. Business 47(2) (1974) 186.

\bibitem{13}
N.Chen, R.Roll, and S.A.Ross,  J. Business 59 (1986) 383.

\bibitem{14}
F.Black, M.Jensen, and M.Scholes, Working Paper (from SSRN), 1972.

\bibitem{15}
E.F.Fama, and K.R.French, J. Financial Econ. 33 (1993) 3.

\bibitem{16}
S.A.Ross, J. Econ. Theory 13 (1976) 343.

\bibitem{17}
D.B.West, Introduction to Graph Theory, Prentice-Hall, Englewood
Cliffs NJ. 1996.

\bibitem{18}
Gower,J.C., and G.J.S.Ross, Appl. Stat. 18(1) (1969) 54.

\bibitem{19}
B.S.Everitt, Cluster Analysis, Heinemann Educational Books,
London, 1974.

\bibitem{20}
J.C.Gower, Biometrika 53 (1966) 325.

\bibitem{21}
J.B.Kruskal, Proc. Am. Math. Soc. 7 (1956) 48.

\bibitem{22}
R.C.Prim, Bell System Techn. J. 36 (1957) 1389.

\bibitem{23}
S. A. Ross, J. Econ. Theory 13 (1976) 343.

\bibitem{24}
H.H.Harman, Modern Factor Analysis, The University of Chicago
Press, Chicago, 1976.

\bibitem{25}
D.N.Gujarati, Basic Economatrics, McGraw-Hill International
Editions, New York, 1998.

\bibitem{26}
H.F.Kaiser, Psychometrika 23 (1958) 187.

\end{thebibliography}
\end{document}